\documentclass[twocolumn,showpacs,preprintnumbers,amsmath,amssymb,pre]{revtex4}
\usepackage{epsfig}

\begin{document}

\title{Quantum Statistics and Thermodynamics in the Harmonic Approximation}
\author{J.~R. Armstrong, N.~T. Zinner, D.~V. Fedorov, and A.~S. Jensen }
\affiliation{ Department of Physics and Astronomy, Aarhus University, 
DK-8000 Aarhus C, Denmark}

\date{\today}

\begin{abstract}
We describe a method to compute thermodynamic quantities in the
harmonic approximation for identical bosons and fermions
in an external confining field.
We use the canonical partition function where only
energies and their degeneracies enter.  The number of states of given
energy and symmetry is found by separating the center of mass motion,
and counting the remaining states of given symmetry and excitation
energy of the relative motion. The oscillator frequencies that enter
the harmonic Hamiltonian can be derived from realistic model parameters
and the method corresponds to an effective interaction approach based on
harmonic interactions. To demonstrate the method, we apply it to systems in two dimensions. 
Numerical calculations are compared to a brute force method that is considerably
more computationally intensive. 
\end{abstract}

\maketitle

\section{Introduction}\label{secintro}
The many-body problem cannot be solved exactly for realistic
interactions and systems.  Numerous approximations have been
formulated and applied over the years. One of the problems is that
many particles require a large Hilbert space to allow for the many
possible correlations. In fact, the space typically grows exponentially
with the particle number. The necessary reduction of the Hilbert space
to obtain a tractable problem 
has to be accompanied by a correponding transformation of the
interaction which in turn has to be used in the smaller space.  This
is the method of effective interactions in reduced spaces.

A different approach starts by an approximation applied to the 
interaction itself. In many physical fields, the two-body interaction
is a complicated entity, and therefore it is desirable to reduce
this complexity. One such direction is to perform a harmonic 
approximation of the interaction Hamiltonian which leaves
all terms (one- and two-body) as a polynomial of at most second order
in the coordinates of all particles. This yields
an exactly solvable model for the many-body system \cite{magda00,arms11}.
However, to utilize such an approach, 
the input parameters must be adjusted to reproduce properties
of realistic systems for low particle numbers. This could be 
two-body binding and scattering information and structural expectation 
values.

This sort of approach has been pursued in nuclear and condensed-matter
physics for many years \cite{bohr74,fetter71} as a source of 
exact insight into many-body problems \cite{sutherland04} when numerics are either 
intractable or hard to interpret. More recently, cold atomic gases have
emerged as an arena and testing ground for various models due to 
the large degree of controllability in experiments \cite{bloch2008}. 
In particular, interaction can be controlled through interatomic
Feshbach resonances \cite{chin10}.
However, these systems are almost exclusively studied in the 
presence of an external confining potential which can often 
be modelled by a (possibly deformed) harmonic oscillator. For 
experimental setups with very deformed traps, the geometry can 
in fact be manipulated to study effective one- or two-dimensional
dynamics. Very recently, 
it has even become possible to explore the few-body limit with 
trapping of just a few atoms near the degenerate regime \cite{jochim10}.

At the moment, the physics of ultracold atomic gases is, however, dominated
by short-range interacting neutral atoms \cite{bloch2008}. In addition, 
since the samples are often extremely dilute, the details of the 
potential at short-range does not matter, and only low-energy scattering
information is important \cite{bethe49}. The replacement of the potential by a 
regularized zero-range potential is therefore often justified. In 
the presence of a harmonic confinement, the solution of the two-body
problem with a zero-range potential was presented by Busch {\it et al.}
\cite{busc98}. This solution was subsequently confirmed in experiments
\cite{stoferle06}. More recently, this model has been used as a 
starting point for addressing the few- and many-body problem in 
a controlled manner in both nuclear and cold atom physics \cite{haxton02}.

The zero-range approximation scheme is one approach to effective 
interactions. Here we pursue a different one via harmonic expansion
of the interaction Hamiltonian. System that can be described in 
some regimes of parameter space by harmonic Hamiltonians are extremely
useful due to their solvability. Of course, as mentioned above, the 
parameters have to be meningfully related to realistic physical 
systems. This can be done through fits to two-body properties 
as described previously in Ref.~\cite{arms11}. Here we take a 
parametric approach and study the details of the harmonic 
approximation for identical quantum particles 
as function of the two relevant frequencies for such a system; 
the (internal) interaction frequency and the (external) trapping
frequencies, both of which are described by (isotropic) harmonic 
terms. Deformation can be easily implemented and will not be 
considered here.

The objective of the present paper is to describe a method for 
calculating thermodynamical quantities for systems of 
identical bosons and fermions, a problem of relevance for 
any subfield of physics concerned with quantum mechanical behavior 
of multiparticle system in equilibrium.
This requires full access to the partition function
and the thermodynamic quantities that can be derived from it.  
The quantum statistics of ensemble of particles is needed here, and 
we present a general method for obtaining such information for harmonic
Hamiltonians that works for any number of particles in principle.

There is great interest in the universal thermodynamics 
of strongly interacting Fermi gases \cite{hu2007}, and 
measurements have determined the equation of state of 
the system \cite{ufermi}. Those studies are mostly concerned
with trap-averaged quantities that are less sensitive to 
sudden changes in thermodynamic behavior as expected near
phase transitions. However, recently it has become possible to 
determine local properties of the gas by density fluctuation
measurements \cite{fluc}. Quantities that involve derivatives
of the energy or pressure like heat capacity and 
compressibility can thus be used to study for 
instance superfluid properties of both fermionic \cite{ku2011} 
and bosonic systems \cite{hung2011}. In addition,
the same experiments are also able to explore 
effects of changing the dimensionality of the system through 
optical lattice potentials.

In the current presentation we focus on extending th harmonic 
approximation into the statictical regime, and more 
particularly on the technicalities of the method used to
achieve this. To demonstrate our 
method we consider the case of a two-dimensional system of 
identical bosons or fermions for different ratios of interaction
to external trapping frequencies. The paper is organized as follows.
In Sec.~\ref{secmethod} we briefly recap the harmonic method and then
discuss the partition
function in
the harmonic approximation. The most important part of the method is the 
counting of the degeneracies in the partition function which we discuss in 
extended detail. The thermodynamic quantities are briefly outlined and a
discussion of how to related the parameters in the harmonic Hamiltonian to the 
model by Busch {\it et al.} \cite{busc98} is given. 
Practical numerical results are obtained
and discussed in Sec.~\ref{secthermo} for a system of cold atoms confined to two
spatial dimensions.  Both bosons and fermions are explicitely
investigated. Finally, give a summary and an outlook in Sec.~\ref{seccon}.

\section{Method}\label{secmethod}
We consider a system of $N$ identical quantum particles, fermions
or bosons, that have
two-body interactions, all of which are confined by an external field. 
All interactions are substituted by carefully chosen harmonic
oscillators. The solutions of the Hamiltonian can then 
be found explicitly and treatment of
thermodynamic properties is possible. 
We first present the necessary ingredients
of the model, and continue to develope the method for calculating the
partition function. This includes the key discussion of how to 
obtain the degeneracies of the many-body states.

\subsection{Formulation of the model}
The $N$-body Hamiltonian contains
kinetic energy, one- and two-body interaction terms, all of which we 
assume can be written in terms of second order polynomials in the 
particle coordinates when proper choice of effective interaction 
parameters have been made. The resulting harmonic
oscillator structure can be separated into terms related to each of
the Cartesian coordinates and the Hamiltonian in one, two, and three
dimensions is a sum of terms from each coordinate. For example, in the
$x$-direction, the effective Hamiltonian is
\begin{eqnarray}
\hat{H}_x&=&-\frac{\hbar^2}{2m}\sum_{k=1}^N\frac{d^2}{dx_k^2}\label{genH}
 +\frac{1}{2} m\sum_{k=1}^N\omega_0^2x_k^2 \\
&+&\frac{1}{4} \mu \sum_{i,k}\omega_\textrm{int}^2(x_i-x_k)^2
 +V_{x0} \;\;, \nonumber
\end{eqnarray}
where $m$ is the mass of the identical particles and $\mu=m/2$ is the reduced mass.
The one-body (external) harmonic trap frequency is $\omega_0$, while
the two-body interaction frequency between all pairs is denoted 
$\omega_\textrm{int}$ (internal). $V_{x0}$ is a constant energy shift which 
is included for generality. Double counting of
interactions account for the factor $1/4$.  
For two and three dimensions, the
total Hamiltonian is $H= H_x + H_y $ and $H= H_x + H_y + H_z$. 
We will assume that the external harmonic trap is isotropic, extension to 
the deformed case are straightforward and will be addressed in future 
work.
All particles are identical and interact in the same way which means that
$\omega_\textrm{int}$ is independent of $i$ and $k$.  Also the one-body
frequency, $\omega_0$, is independent of $k$.

The Schr\"{o}dinger equation for the Hamiltonian in Eq.~\eqref{genH}
can be solved analytically for any particle number (details can be found
in \cite{arms11}) The pertinent
properties in the present context are that the energies, $E_j$, of the
excited states have oscillator character. Here $E_j$ denotes the 
$j$th {\it many-body} excited state, i.e. the energy of all $N$
particles when they are distributed in a certain way in the oscillator
levels that come out of the diagonalization of the full Hamiltonian. 
In this paper we will concentrate mostly on the case of a two-dimensional
system, and we therefore specialize to this case now. The energy of 
a many-body state, $E_j$, can be written
\begin{eqnarray} 
 && E_j = \hbar\omega_r (N_\textrm{rel} + N-1) +
 \hbar\omega_0  (N_\textrm{cm} + 1) \;,  \label{ej}  \\ \label{e30}  
 && N_\textrm{rel} =  \sum_{k=1}^{N-1} (n_{x}^{k}+ n_{y}^{k}) \;,\\     
 &&N_\textrm{cm} =  n_{x}^\textrm{cm} + n_{y}^\textrm{cm} \;,\label{e31}
\end{eqnarray}
where $n_{x}^{k}$ and $n_{x}^\textrm{cm}$ are quanta in the $x$-direction
related to the relative and center-of-mass excitations with the former
carrying the index $k$ running over all $N$ particles (and similar terms
for the $y$-direction). The total number of relative and center-of-mass 
quanta in the many-body state $j$ are denoted $N_\textrm{rel}$ and $N_\textrm{cm}$
respectively. Note that $n_{x}^{k}$, $n_{x}^\textrm{cm}$,
$n_{y}^{k}$, and $n_{y}^\textrm{cm}$ must all be specified to completely
characterize the many-body state of energy $E_j$. However, more than one set
of individual quantum numbers can lead to the same energy $E_j$ and this 
determines the degeneracy of the state, which we denote $g_j$ and discuss
at length below. $E_j$ and $g_j$ are all that is needed to compute the 
partition function and thus the thermodynamic properties of the system.

The constants, $\hbar\omega_r(N-1)$ and $\hbar\omega_0$ in the 
formula for $E_j$,
are from the two-dimensional zero-point energy of intrinsic and center
of mass parts, respectively.  
The one-body frequency, $\omega_0$, 
correspondings to center of mass motion.  The
other frequency, $\omega_r$, is $N-1$ times degenerate for
identical particles.  It is given by the relation \cite{arms11}
\begin{eqnarray} \label{e40}
\omega_r^2 = \omega_\textrm{int}^2 \frac{N}{2}  +  \omega_0^2  \;.
\end{eqnarray}
If the two-body interaction goes to zero ($\omega_\textrm{int}\to 0$), 
$\omega_r\to\omega_0$, and the resulting eigenvalues and
eigenfunctions converge to those of the external one-body 
trapping potential.
The solutions are $N$ non-interacting identical particles
occupying $N$ states in a harmonic oscillator.  For interacting
particles, we obtain instead $N-1$ identical frequencies with 
value given by Eq.~\eqref{e40}. The extension of the above discuss 
to three dimensions involves also the $z$-direction and is 
straightforward, although the extra quantum numbers makes
the degeneracy larger and the counting of many-body states
more involved.

With the harmonic Hamiltonian, the exact spectrum of excited many-body 
states, $E_j$, are thus available for a
given number of particles. For the thermodynamic calculations, the
natural choice is to use the canonical ensemble applicable for a
definite number of particles. The canonical partition function is
\begin{equation}
  Z = \sum_j g_j \exp(-E_j/T),
\label{genpartfcn}
\end{equation}
where $g_j$ is the degeneracy of the $j$th many-body state and $T$ is
the temperature measured in energy units (the Boltzmann constant, $k_B
=1$).  Once $Z$ is obtained, all thermodynamic quantities can be
derived from it. The most involved question is how to obtain $g_j$ 
for a given energy of the many-body system. This will now be discussed.

\subsection{Symmetry constraints on degeneracies}
The only quantities in the partition function are the energies, $E_j$, and
their degeneracies, $g_j$.  
The energies are appropriate multiples of the
harmonic oscillator frequencies introduced above.  
The degeneracies depend
strongly on which symmetry restriction is imposed.  Problems
arise when all, or a group of, particles cannot be
distinguished because they are identical and allow occupation
of the same states.  We then have to count the number of (anti-)symmetric
states for the indistinguishable particles of a given excitation
energy, $E_j$, and the corresponding degeneracies, $g_j$, of each
{\it many-body} state of energy $E_j$. 
We now describe our procedure to obtain $g_j$. It is
based on knowledge of the non-interacting system, a recursive 
procedure, and the fact that when the two-body interaction
vanishes, the
degeneracies remain the same as we will now explain.

We start with $N$ identical particles without any two-body
interaction ($\omega_\textrm{int}\to 0$), 
and all in the same external trap with frequency
$\omega_0$. This Hamiltonian will have $N$ degenerate solutions
of frequency $\omega_0$ per spatial dimension (so $2N$ for 
the two dimensional case).
The many-body solutions are
the products of $N$ combinations of the single-particle harmonic
oscillator wave functions.
For identical fermions, the Pauli principle, or equivalently the
antisymmetrization, requires that each of the single-particle states
at most is occupied by one particle.  In addition, if this condition
of occupancy by at most one particle for each state is fulfilled, then
one and only one antisymmetric state is uniquely constructed as the
Slater determinant of the $N$ singly occupied single-particle states.

Importantly, the energy can be characterized by the sum, 
$N_\textrm{tot}=N_\textrm{rel}+N_\textrm{cm}$, of
single-particle oscillator quantum numbers for all particles in the 
state $E_j$,
as seen in Eqs.~\eqref{e30} and \eqref{e31}.
Assume that we have
computed the number, $g_\textrm{non}(N_\textrm{tot})$, of properly 
symmetrized states of
non-interacting identical particles in the external potential.
The quantity $g_\textrm{non}(N_\textrm{tot})$ can be computed
in a simple manner by trail-and-error, i.e. taking all possible
permutations and testing which are completely symmetric and 
which are completely antisymmetric. This is, however,
notoriously 
difficult since the computational effort
is exponentially increasing with particle number. Also,
it does not distinguish between center of mass motion and 
relative motion and this needs to be disentangled. This can
be done by recalling the symmetry of the center of mass
as we now discuss.

The center of mass motion is a symmetric mode in all permutations of the
particle coordinates, simply because the center of mass coordinate is
the sum of the individual particle coordinates.  This motion is always
symmetric, independent of the number of quanta in this mode, $N_\textrm{cm}$.
Therefore, the relative motion determines the symmetry of the total
wave function.  The excited states consist of a combination of two
distinctly different types of excitations. They are characterized by the
number of quanta in the center of mass motion, $N_\textrm{cm}$, and the
number of quanta in the relative motion, $N_\textrm{rel}$.

We denote the number of properly (anti)symmetric $N$-body states by $g_{p}(N_\textrm{rel})$,
which is the number of states that have the proper symmetry in terms of 
the {\it relative} coordinates. 
It can be
computed from knowledge of the total number of non-interacting
many-body states, $g_\textrm{non}(N_\textrm{rel})$, for all values of
$N_\textrm{tot}=N_\textrm{rel}+N_\textrm{cm}$.
To do so, one must subtract
the number of states of different center of
mass quanta from the total number of non-interacting many-body states 
in a recursive manner, i.e.
\begin{equation} \label{e103}
g_{p}(N_\textrm{tot}) =  g_\textrm{non}(N_\textrm{rel}) - 
 \sum_{N_\textrm{cm}=1}^{N_\textrm{rel}}   d_{N_\textrm{cm}} g_{p}(N_\textrm{rel}-N_\textrm{cm})\;,
\end{equation}
where the degeneracy, $d_{N_\textrm{cm}}$, of the
center of mass state depends on the spatial dimension of the
oscillator; $d_{N_\textrm{cm}}=1$ for one, 
$d_{N_\textrm{cm}}=N_\textrm{cm}+1$ for two, and 
$d_{N_\textrm{cm}}=(N_\textrm{cm}+1)(N_\textrm{cm}+2)/2$ 
for three dimensions, respectively. The formula comes from the 
fact that we know only the total number of quanta, $N_\textrm{tot}$, 
but {\it not} $N_\textrm{rel}$ and $N_\textrm{cm}$ individually. We need
to recursively subtract the states that correspond to 
$N_\textrm{cm}=1$, 2, $\ldots$, $N_\textrm{tot}$ as the formula
prescribes. The factor $d_{N_\textrm{cm}}$ is due to the degeneracy
of the center of mass motion itself for given number of quanta, $N_\textrm{cm}$.
Note that $g_{p}(0)$ is well-defined
and corresponds to a state with no quanta of excitation, i.e. the 
ground state. It can be zero or non-zero depending on $N$ and whether
the particles are bosons or fermions.
Eq.~\eqref{e103} determines the degeneracy of an $N$-body 
state in a harmonic system and is one of the main results 
of this work.

The degeneracy can now be calculated iteratively from
Eq.~\eqref{e103} for any number of quanta, $N_\textrm{tot}$
from a starting point based on a non-interacting system 
of identical particles in a trap.
These states must then be either completely symmetric or
antisymmetric with respect to interchange of all coordinates of any
pair of particles to obey either bosonic or fermionic statistics.  
Now we add the two-body interaction on top of the
external one-body potential. The external frequency, $\omega_0$, remains
in the spectrum
corresponding to the center of mass motion, and an
$N-1$ times degenerate frequency, $\omega_r$, appears corresponding to
relative motion. The absolutely crucial point is that 
the degeneracies do {\it not} depend on the
two-body interaction strength which will only influence the value of the
degenerate frequency.  The counting remains completely unchanged,
because the number of states of a given symmetry is a discrete number
which does not change with continuous variation of the interaction.

Mathematically, this corresponds to a continuous map (the scheme 
for counting degeneracies) from a continuous interval (the interaction
frequency, $\omega_r$) to a discrete set (number of states, $g_j$, 
with given energy). Such a map must necessarily be a constant map.
The underlying point here is that our splitting into $N_\textrm{cm}$
and $N_\textrm{rel}$ is done in a way that moves the states that 
are related to the relative motion when $\omega_r$ is increased 
from zero to its full value, while the states that are related to 
center of mass motion remain constant in energy since $\omega_0$ 
remains a solutions.

We confirmed this simple counting method by an elaborate 
and computationally very slow brute force
procedure which is much worse than computing $g_\textrm{non}(N_\textrm{tot})$
above. The (anti-)symmetrization of each of the computed
wave functions is performed by permutations of all the coordinates
followed by extraction of a basis with a size equal to the smallest
number of linearly independent states.  For sufficiently small $N$ 
this can be computed in reasonable time and comparison can be made. 
In all cases considered below we found perfect agreement between 
the numerics and the procedure outlined above. The procedure
above is, however, far superior since it needs only counting 
of states with given energy, {\it not} the full wave functions.

\subsection{Thermodynamic quantities}
We can now use Eq.~\eqref{e103} to get the number of center of mass
states and the number of relative states as functions of the sum of
corresponding quanta.  Along with the frequencies, this is all we need
to calculate the partition function in Eq.~\eqref{genpartfcn}. This 
simplification occurs since the structure of the wave
functions do not enter in the partition function.

The center of mass mode of frequency $\omega_0$ separates in
Eq.~\eqref{genpartfcn} because both the exponential function and the
degeneracy factorizes, that is $g_j = g_{as}(N_{rel}) d_{N_{cm}}$,
where the center of mass contribution is
analytical and given below Eq.~\eqref{e103}. We have
\begin{eqnarray}  \label{idpartfcn} 
Z(N,T)&=&  Z_{cm}(N,T) \times Z_{rel}(N,T), \\ \nonumber 
 Z_{cm}(N,T) &=& \left(\frac{\exp(-E_{gs}/(2T))}{1-\exp(-E_0/T)} \right)^D,
  \\ \nonumber
 Z_{rel}(N,T) &=& \sum_{N_{rel}=0} g_{as}(N_{rel}) \exp(-N_{rel} \hbar \omega_r/T) \;,
\end{eqnarray}
where $D$ is the dimension, $E_0 = \hbar \omega_0$, and $E_{gs}= E_0
+ (N-1) \hbar \omega_r + V_0$. In practice, the infinite sum over relative quanta has
to be limited by a cut-off value which is chosen sufficiently
high to achieve the required accuracy.  Certain quantities are more 
sensitive to the cut-off than others, such as those with more derivatives 
of the partition function (heat capacity, compressibility, etc.). For 
all the results presented below we have checked convergence by 
increasing the cut-off and identifying the value of $T$ below 
which the relevant quantities remain unchanged.
From the canonical partition function, Eqs.~\eqref{genpartfcn} or
\eqref{idpartfcn}, we get immediately the basic quantities of
energy, $E$ and free energy, $F=E-TS$, that is
\begin{eqnarray} \label{e113}
F = - T \ln Z \;,\;\; 
E = T^2 \frac{\partial \ln Z}{\partial T} \;,\;\;
C_V= \frac{\partial E}{\partial T} \;,
\end{eqnarray}
where $S$ is the entropy, and $C_V$ is the heat capacity \cite{mcqu76,huan87}.

In thermodynamic formulations of macroscopic systems, both temperature
and volume is usually external parameters. Both the energy, $E$,
and heat capacity, $C_V$, defined above, are in principle the derivatives
for a fixed volume, $V$ (and also fixed $T$ and $N$). 
In our case with an externally confined 
$N$-body system the volume is not the fixed quantity, but rather 
it is the external trapping frequency, $\omega_0$, that is 
fixed. The derivatives above are thus taken for fixed $\omega_0$.
Usually, variation of the volume allows information about pressure and
compressibility. Definitions of these quantities and connection to
their thermodynamic counterparts require a precise translation between
$\omega_0$ and volume. Here we employ the simplest choice
and define the volume via the length
parameter related to the external trap, $b$, which is defined by 
$b^2=\tfrac{\hbar}{m\omega_0}$. We now have $V=s_D b^D$, where 
$s_2=\pi$ and $s_3=4\pi/3 $ are the two and three dimensional 
angular surface areas. Derivatives of any quantity, $Q$, 
with respect to volume can now be written
\begin{equation} \label{e127}
 \frac{\partial Q}{\partial V} =  \frac{\partial Q}{\partial \omega_0} 
  \frac{\partial \omega_0}{\partial V} =
 - \frac{2\omega_0}{DV} \frac{\partial Q}{\partial \omega_0}.
\end{equation}
For the pressure, $P$, we find
\begin{eqnarray} \label{e137}
P =-\frac{\partial F}{\partial V}=\frac{2\omega_0}{DV} \frac{\partial F}{\partial \omega_0},
\end{eqnarray}
which, from the definition in Eq.~\eqref{e113}, gives 
\begin{eqnarray} \label{e147} 
 &P = P_{cm} + P_{rel},&   \\  
&P_{cm} = \frac{2T\omega_0}{DV}\frac{\partial \ln Z_{cm}}{\partial \omega_0},&\\
&P_{rel} = \frac{2T\omega_0}{DV}\frac{\partial \ln Z_{rel}}{\partial \omega_0}.&
\end{eqnarray}
The derivatives are easily worked out since they only contain
$\partial \omega_r /\partial \omega_0 = \omega_0/\omega_r$.

Continuation to the second derivative with respect to the volume
produce the isothermal bulk modulus, $B_T$, defined as
\begin{eqnarray} \label{e157}
B_T=1/\kappa_T=-V \frac{\partial P}{\partial V} 
= \frac{2\omega_0}{D} \frac{\partial P}{\partial \omega_0}  \; ,
\end{eqnarray}
which is the reciprocal of the isothermal compressibility, $\kappa_T$.
Again two terms arise related to center of mass and relative degrees
of freedom.  Several terms appear by carrying out the two
derivatives.  This kind of compressibility, that of a response of a system 
to an applied pressure, is different to the compressibility of a self-bound 
system such as a nucleus \cite{wong98}. 
For self-bound systems, the compressibility refers
to resistance to density fluctuations. This can be written as the 
second derivate of the energy per particle with respect to the Fermi momentum
for fermionic systems.

\begin{figure}
\includegraphics[width=0.48\textwidth]{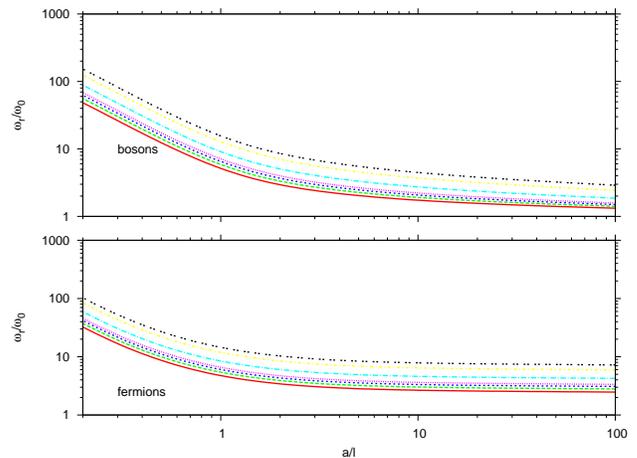}
\caption{The ratio between $\omega_r$ from Eq.~\eqref{e40} and the external trap frequency, $\omega_0$,
  as function of the scattering length, $a$, in two dimensions for the
  Busch model \cite{busc98} with $s$-wave interactions for bosons (upper) and for the $p$-wave interaction 
  identical (i.e. spin-polarized) fermions (lower).  
  In both plots, the different curves correspond to different particle numbers.  
  From the bottom to top, the curves are for 3, 4, 5, 6, 10, 20, and 30 particles. The 
  length parameter is the trapping length, $l=b$, introduces in the text.}
\label{busch}
\end{figure}

Before we present results for degeneracies and various thermodynamic
properties, we discuss how one can fix the 
two-body interaction term to capture the properties of the 
realistic system which have interactions that are not harmonic. In the 
case of identical bosons, the procedure was already presented in 
Ref.~\cite{arms11} but we repeat the arguments here for completeness.

We assume that we are in the situation that the real two-body 
interaction is of much shorter range than the external trap length $b$
above. In this case, we can use the zero-range model of Busch {\it et al.}
\cite{busc98} to obtain the solutions in the trap.
The effective two-body interaction that we want to use
is a harmonic oscillator, which in general contains two parts;
an oscillator frequency, $\omega_\textrm{int}$, and a shift, $V_0$. 
To fix these parameters we employ the conditions that the two-body
harmonic interaction should reproduce the correct two-body binding
energy of the ground state in the Busch model, and also the 
spatial extend of the wave function. The latter condition is 
implemented by calculating the root-mean-square radius, 
$\langle r^2\rangle$, in the exact solution and fixing the 
oscillator parameter to reproduce value. The energy shift is 
suqsequently tuned so as to also reproduce the two-body 
binding energy.

Clearly, since the Busch model uses a zero-range interaction, 
the only scale left is the external trap frequency, $\omega_0$. 
Therefore what we obtain from this is the quantity $\omega_\textrm{int}/\omega_0$.
We ignore the shift from now on since it merely provides an
overall shift in the $N$-body harmonic Hamiltonian which is not 
of interest in the current paper. The next step is to 
insert this value into Eq.~\eqref{e40} to obtain $\omega_r/\omega_0$,
which will now naturally depend on $N$. In the upper panel 
in Fig.~\ref{busch} we show the result of this procedure for 
the case of bosons in two dimensions (more detail for bosons
in both two and three dimensions can be found in Ref.~\cite{arms11}).
The results are plotted as function of the two-dimensional 
scattering length $a$. Note that when $a$ is small, the 
ground state becomes strongly bound, while for large $a$ 
it goes to the non-interacting limit. This is clearly
reflected in the behavior of $\omega_r/\omega_0$.

The result of this procedure agrees rather well with predictions from the
well-studied problem of bosons interacting through a zero-range 
interaction \cite{lovelace87} in the strong and weak coupling 
limits \cite{arms11}, also in the case where higher-order interaction
terms are included \cite{fu03}. Recently, a very similar procedure has 
been used to study polar molecules in layered system \cite{armstrong2010}
where excellent agreement with exact methods has been found both 
for isotropic \cite{armstrong2011} and anisotropic potentials \cite{volosniev2011}
which have reasonably large potential pockets. Incidentally,
one-dimensional dipolar system also turn out to have such
potential pockets \cite{wunsch2011} and applying the harmonic
approximation to these systems is an interesting direction 
for future research.

For identical (spin-polarized) fermions a complication arises
from the fact that the original Busch model applies to 
particles interacting through an isotropic $s$-wave interaction 
only. The Pauli principle dictates that $s$-wave interactions
are zero for identical fermions, and therefore the interactions
must have $p$-wave (or higher odd partial wave) character.
In two dimensions, the equivalent of the Busch model with 
$p$-waves can be solved and the spectrum turns out to be 
very similar to the $s$-wave case except for (unimportant) 
shifts \cite{pwave2011}. However, a further complication 
arises since the corresponding ground state wave function 
is singular at the origin in a manner that does not allow
it to be normalized \cite{kanjilal2006}. This can be fixed
by a properly defined scalar product as discussed for the 
three-dimensional case in Ref.~\cite{pri2006}. We will not
pursue this approach here but instead we note that any
higher order structural average of the type $\langle r^n \rangle$
with $n$ an integer is still perfectly convergent.
We thus consider a higher order average, 
$\langle r^4\rangle/\langle r^2\rangle$, in order to fix the 
oscillator parameter for identical fermions. We are 
thus tacitly using that the $p$-wave spectrum is 
very similar to the $s$-wave one and assume that 
the wave function should be so as well. Of course, it 
must be kept in mind that this is {\it not} at odds
with the Pauli principle since the two-body wave function 
in the $p$-wave channel is still zero at the origin. In any case,
the Pauli principle is exactly enforced on the $N$-body system
as discussed above. 

The results for $\omega_r/\omega_0$ are
shown in the lower panel of Fig.~\ref{busch}. The 
similarity to $s$-waves and bosons is quite clear, and the only
difference seems to be slightly lower overall values for the 
fermionic case. Below we will illustrate our method by 
calculating thermodynamic quantities for both fermions and 
bosons. Here the only thing that matters is the ratio $\omega_r/\omega_0$.
One can then reverse the process and use Fig.~\ref{busch} to 
obtain a corresponding value of the scattering length and 
thus compare to a realistic system with trapped bosons or
single-component fermions. Since this is mainly a discussion of 
method, we will not dwell on this any further. An important 
thing to note, however, is that since the ratios $\omega_r/\omega_0$
are similar for bosons and fermions, the results below will 
truly isolate the statistical behavior coming from the 
exchange requirements in the different cases.

\section{Thermodynamic Results}\label{secthermo}
We now proceed to demonstrate our methods by computing 
the density of states and the thermodynamic quantities that
have been introduced above. They depend
on particle mass, dimension, number of particles and their quantum
statistics, temperature, one- and two-body frequencies.  If we measure
all energies in units of $\hbar \omega_0$, all results for given
particles in $D$ dimensions, depend only on the two ratios, $T/\hbar
\omega_0$ and $\omega_r/\omega_0$.  
This implies that any kind of
interaction model used to define the effective harmonic hamiltonian
only has to provide these two quantities as pointed out above.

\begin{figure}
\includegraphics[width=0.48\textwidth]{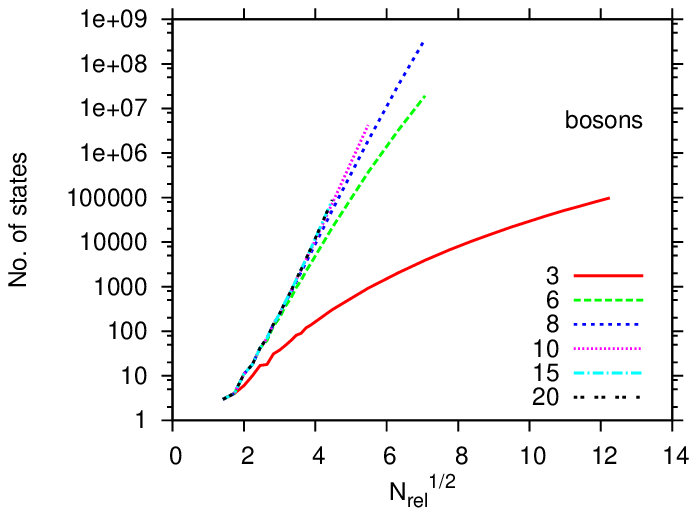}
\includegraphics[width=0.48\textwidth]{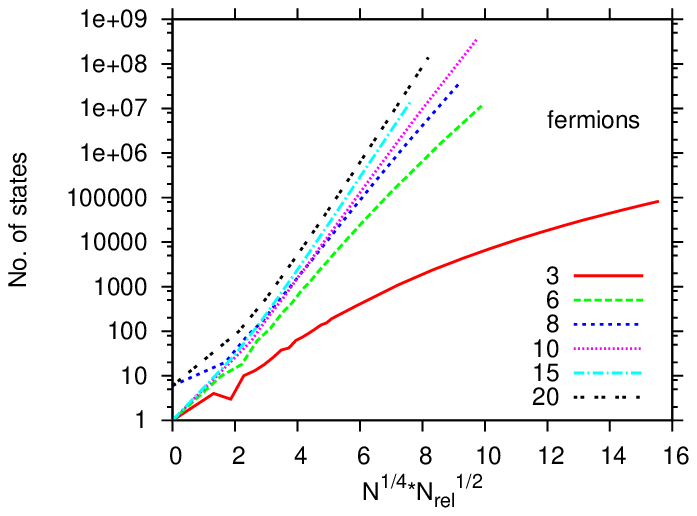}
\caption{The number of completely symmetric (bosons) and antisymmetric (fermions) states as function
  of excitation energy for different numbers of particles in a two-dimensional harmonically trapped system. 
  The particle numbers are indicated on the plots. Notice that the horizontal axis for bosons is $N_\textrm{rel}^{1/2}$,
  while it is $N^{1/4}N_\textrm{rel}^{1/2}$ for fermions.}
\label{fermbos}
\end{figure}

\subsection{Degeneracies}
First we discuss the degeneracies themselves which are found by the
procedure discussed above, essentially through the recursive
formula Eq.~\eqref{e103}.
While the number of non-interacting (anti-)symmetric
states increases exponentially as function of excitation energy,
the
numbers can, however, be tremendously reduced by the symmetry requirements.  
A simple example indicating this tendency is that for 
bosons all particles are allowed to occupy the lowest level with zero
quanta, $N_{tot}=0$. In contrast, a given number of fermions has a
minimum sum of quanta, $N_{gs}$, in the ground state, that is $N_{tot}
\ge N_{gs} \neq 0$.  We show the numbers of given symmetry as
functions of excitation energy in Fig.~\ref{fermbos}. 

Since we are dealing with harmonic oscillators, 
a given energy corresponds to a given total number of quanta.  
The
starting point for the ground state 
is naturally $N_\textrm{cm}=0$ and $N_\textrm{tot}=N_\textrm{rel}$, which
is larger for fermions than for bosons due to the requirement that
at most one fermion can occupy each state.
The corresponding higher degeneracy implies therefore that the number
of states of given excitation energy is larger for fermions than for
bosons because the latter do not need to obey the Pauli Principle.  
For bosons, the lowest
excited state consisting of one quantum added to zero quanta in the ground state 
can only be a center of mass excitation, since one quantum in the relative
motion would be antisymmetric in one pair of coordinates 
(it would have to be a state of $p$-wave/parity-odd symmetry).  
The lowest
completely symmetric excited state of the relative motion appear
for $N_\textrm{rel}=2$. In conclusion, there is always a gap in the excitation spectrum for
bosons corresponding to $2\hbar\omega_r$ where fermions only have
a gap of $\hbar\omega_r$. This will be reflected in the
temperature dependence below.

For bosons (upper part of Fig.~\ref{fermbos}) the same degeneracy is
found for low energy (small total number of quanta) for all particle numbers.
In the upper plot of Fig.~\ref{fermbos} the horizontal axis is 
$N_{\textrm{rel}}^{1/2}$ since we have found that this is a good measure
for the excitation energy in the system. 
For fermions, one can give 
a qualitative argument for the exponential behavior which we discuss below.
The similarity of degeneracies at low energy for bosons can be understood 
by the direct counting procedure where a given
number of particles, $N$, has to be distributed to add up to the total
number of quanta, $N_\textrm{rel}$. 
First, $N-1$ particles are placed in
the lowest oscillator level, and the one remaining particle then has
to be placed in the level with total number of quanta equal to
$N_\textrm{rel}$. Then we move the single particle one step down to
$N_\textrm{rel}-1$, and simultaneously compensating by moving one particle
one step up from the lowest level. We continue with these combinations
until we have $N_\textrm{rel}$ particles in the second oscillator level and
all others in the lowest level.  Adding one particle and repeating the
counting process we realize that their is a one-to-one correspondence
between the configurations of $N$ and $N+1$ particles. Going from one to
the other is simply by removing or adding one particle in the lowest
oscillator level with zero contribution to the total number of
quanta. 

For $N$ bosons, the deviation from this universal curve starts for
$N_\textrm{rel}=N+1$. The reason is again found by following the counting
procedure. The configurations with all particles in the lowest
two levels are only possible when $N_\textrm{rel}\le N $. This implies that
we find fewer states when $N_\textrm{rel}\ge N+1 $ for $N$ than for $N+1$
particles. The curves break away from the universal curve for
increasing $N$ when $N_\textrm{rel}=N+1$. 

The degeneracies for fermions have very different behavior, as seen in
the lower part of Fig.~\ref{fermbos}. Notice that the 
horizontal axis is different in the two plots.
We notice a regime of linear dependence which has the same origin as
the exponential square root dependence of excitation energy of the
free Fermi gas level density \cite{bohr74}. 
For two dimensions, the particle number
dependence is roughly $N^{1/4}$ as reflected on the axis in the 
fermion plot. This holds for intemediate excitation energies and for 
relatively large particle numbers. A qualitative understanding 
of the behavior can be obtained in a manner following 
Ref.~\cite{bohr74}. The 
density of {\it single-partile} states at the Fermi level in 
a two-dimensional harmonic trap is roughly
$g_F\sim (2N)^{1/2}/\hbar\omega_0$, while the typical excitation 
energy in the system is $E^*=\hbar\omega_0 N_\textrm{rel}$. 
The {\it many-body} level density in this situation is then 
proportional to
$\textrm{exp}\left[\sqrt{\pi^2 (2N)^{1/2} N_{rel}/6 }\right]$. This explains the 
choice of horizontal axis for fermions in Fig.~\ref{fermbos}.
The suggested linear dependence is not
very clear but the assumptions are not well fulfilled. The 
requirements is that
the excitation energy has to be sufficiently large to allow
statistcial treatment and sufficiently small not to exhaust particles
at the bottom of the potential. This is not true 
for the relatively small particle numbers that we must necessarily
work with to make the problem tractable by both brute force
symmetrization and the counting scheme developed here.

\begin{figure}
\includegraphics[width=0.5\textwidth]{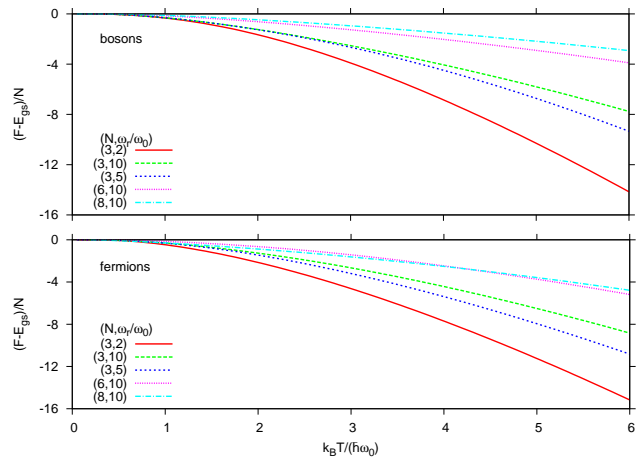}
\caption{Helmholtz free energies of a two-dimensional system, divided by particle number, $N$, for
  several boson (upper) and fermion (lower) systems, identified by the
  number of particles and frequency ratio, $(N,\omega_r/\omega_0)$.
  The ground state energy is subtracted, and the energy unit is $\hbar
  \omega_0$ for both energies and temperatures. The ground state energy, $E_\textrm{gs}$, has been substracted.}
\label{bosF}
\end{figure}

\subsection{Energy and free energy}
We are now in a position to explore the thermodynamics
and to compare bosonic and fermionic
behavior in detail.
The free energies per particle for a two-dimensional system
are shown in Fig.~\ref{bosF} as function of temperature for different
interactions and particle numbers, both for bosons (upper panel)
and fermions (lower panel). Note that the ground state energy
has been subtracted as it is not of interest here.
The curves start out very flat at
zero showing that the low-temperature dependence is at least quadratic.
The decrease with increasing temperature is actually rather similar to
a quadratic behavior with $T$. The behavoir is modified when 
$T$ exceeds $\hbar\omega_0$ and the center of mass mode becomes
active. Notice that the dependence on
particle number is relatively weak.  This is in contrast to the
interaction dependence where the free energy varies substantially more
for the smallest two-body interactions. In addition, we note that 
fermion free energies are always smaller than boson free energies.
This demonstrates that the free energy is entropy driven, i.e. the 
entropy term grows faster than the energy with temperature. Again 
the larger degeneracy of fermionic systems is seen by the lower 
free energies.

\begin{figure}
\includegraphics[width=0.48\textwidth]{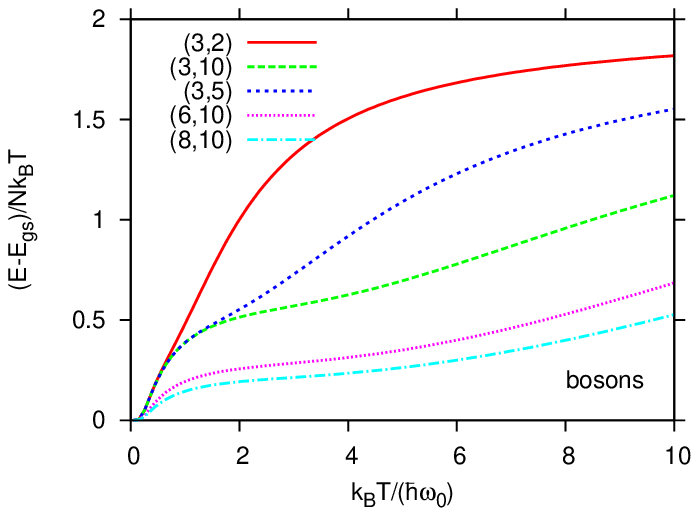}
\includegraphics[width=0.48\textwidth]{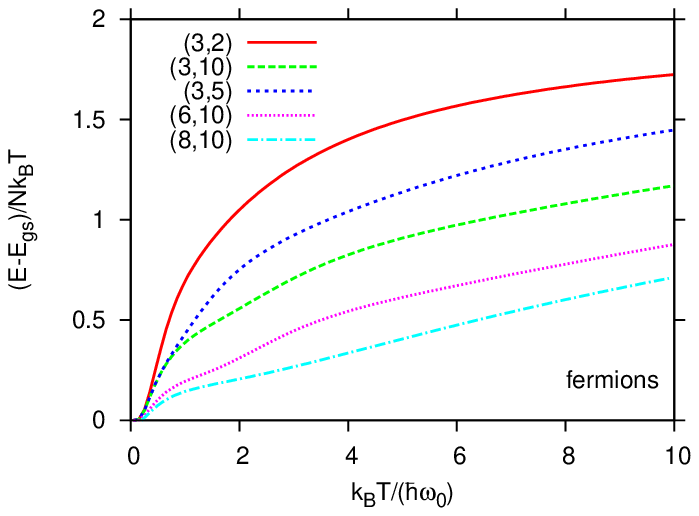}
\caption{ Energies in two dimensions, divided by $Nk_BT$, for
  several boson (upper) and fermion (lower) systems, identified by the
  number of particles and frequency ratios,
  $(N,\omega_r/\omega_0)$.  The ground state energy, $E_\textrm{gs}$, has been subtracted from each curve.}
\label{bosU}
\end{figure}

Fig.~\ref{bosU} displays the energy per particle for bosons and
fermions.  We measure energy in units of $Nk_B T$ to isolate
the high temperature behavior which is $2Nk_B T$ for two 
dimensionsal harmonic system by the equipartition theorem. Again
we subtract the uninteresting constant ground state energy.
The increase from zero at zero temperature is rather steep for both
types of particles. All curves continue to increase with temperature
but much slower after a few units of $\hbar\omega_0$. The
largest energy per $Nk_BT$ is for the system with the least particles and
the smallest frequency ratio. This can be understood from the fact that
for large $\omega_r/\omega_0$, only the center of mass modes
(twice degenerate in two dimensions) are active. However, since
we divide by $N$, the values become smaller for larger $N$, but the
behavior remains the same (self-similar lines for $\omega_r/\omega_0=10$
in Fig.~\ref{bosU}). This is similar for both bosons and fermions. 
We also observe that the equipartition limit 
at high temperature is reached for substantially larger 
temperatures, a clear sign of the interaction effects.

The low-temperature behavior is, however, different for bosons and fermions.
The low-energy $2\hbar\omega_r$ gap in the boson spectra arises due to a ground state
with all particles with zero oscillator quanta of excitation.  Then
there are no states with one quanta of excitation.  For fermions, 
there are only $1 \hbar\omega_r$ gaps.  Therefore the energy exhibits flat
regions at relatively low temperature, but these features are observed 
first for fermions and later for bosons due to the difference in energy gap 
which is related to the activation energy. This demonstrates a clear 
signature of quantum statistics in the harmonic model, but which 
should be expected in generic interacting systems with identical 
particles.

\begin{figure}
\includegraphics[width=0.48\textwidth]{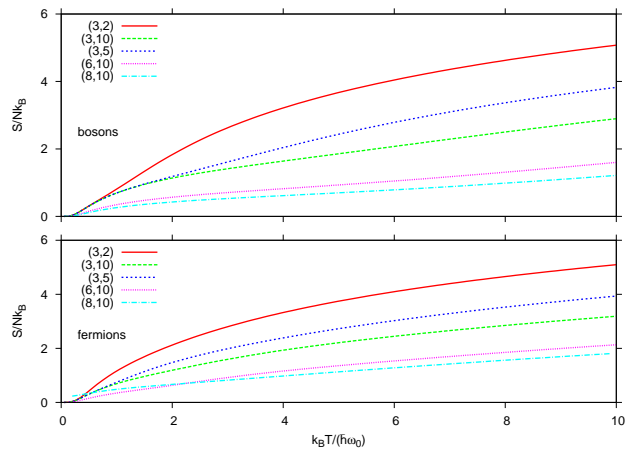}
\caption{ Entropies in two dimensions, divided by $Nk_B$, for boson
  (upper) and fermion (lower) systems, identified by the
  number of particles and frequency ratios,
  $(N,\omega_r/\omega_0)$.  }
\label{bosS}
\end{figure}

Next, we show the entropy, $S$, in Fig.~\ref{bosS}.
We note the increase of available states
per particle as function of temperature goes from zero to values
around a few times the temperature in units of $\hbar\omega_0$.  
For lower ratios $\omega_r/\omega_0$, 
the increase with temperature is faster since
less energy is required to excite the internal modes of the system 
and more states are available, correspondingly increasing $S$.
Again, for larger values of $\omega_r/\omega_0$, the center of
mass is the only active mode at low temperature and the division
by $N$ in the plots explains the lower value of $S$ for larger 
$N$. Only at substantially larger temperatures will both
internal and center of mass modes become active.
The effect of symmetry is not very
pronounced, although it is noticeable that the entropy is larger for
fermionic than for bosonic systems due to the larger degeneracy.

An interesting feature of the fermion plot is that 
for $N=8$ the limit of $S/N$ for $T\to 0$ is non-vanishing.
The number of
available states is finite reflecting that the ground state itself is
degenerate.  This is seen by simply counting of oscillator degeneracy
for a two-dimensional harmonic oscillator.
The lowest three quantum levels can hold 6 identical
fermions (1, 2, and 3, respectively), and the fourth can hold additionally four particles. This
means that eight particles only occupy half of the last level, leaving
the ground state as six times degenerate.  This beavior at zero
temperature then nicely indicate the presence of shell structure.

\begin{figure}
\includegraphics[width=0.48\textwidth]{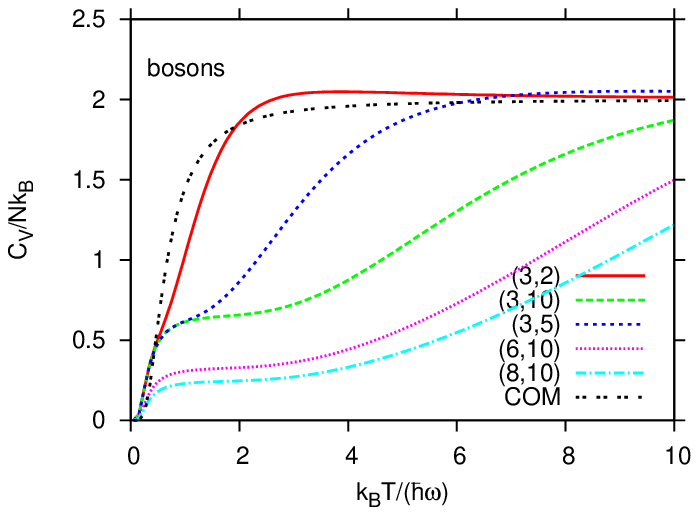}
\includegraphics[width=0.48\textwidth]{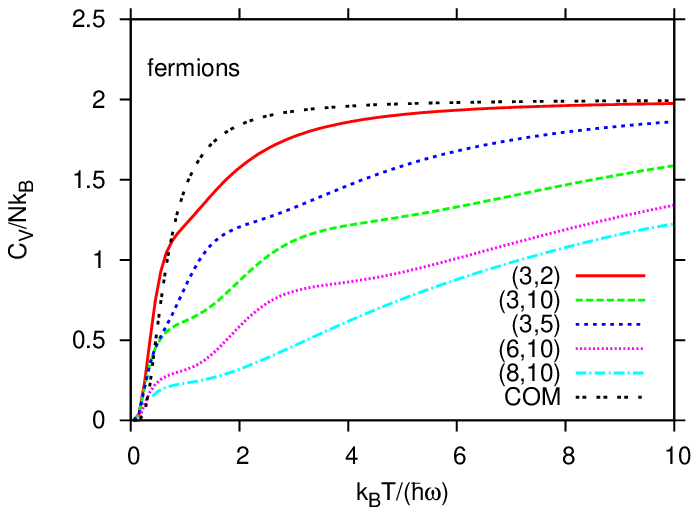}
\caption{ Heat capacities divided by $Nk_B$ in two dimensions for
  boson (upper) and fermions (lower) systems, identified by the number of particles and frequency
  ratios, $(N,\omega_r/\omega_0)$.  The black line shows the
  unscaled value for the mean field center of mass mode.}
\label{bosCV}
\end{figure}

\subsection{Heat capacity and compressibility}
We now consider some derived thermodynamic quantities that 
are of experimental interest in many fields of physics.
The heat capacity and the compressiblity are two such quantities
that can be obtained from second derivatives of the 
partition function as discussed above.
In Fig.~\ref{bosCV}, we
show the heat capacity per particle, $C_V/N$, defined in Eq.\eqref{e113}
as function of temperature for both bosons and fermions.  They both
start with an initial activation of the external trap mode, since the
degenerate relative degrees of freedom all require higher temperature
to be excited.
After a delay, these internal modes are activated, and the heat
capacity increases with temperature.  
The delay and the rate of
increase depend strongly on $\omega_r/\omega_0$ with a slower increase
of $C_V$ for larger interaction ratios.

The tendency to increase slower and in steps 
is related to gaps in the energy
spectrum.  At high excitation energy, the spectrum becomes denser, and
gap sizes larger than the temperature cannot appear.
However, the presence of a gap in the
low-energy spectrum is important for the heat capacity in general.
In fact, this is clearly seen by the fact that bosons have a flat profile 
for a region of low temperature, while the fermions have two
flat plateaus. 
Fermions rise faster initially, and always approach the equipartition
heat capacity from below ($C_V/N\to 2$ for two dimensions).  
The larger gap causes a delay in the
boson systems. The heat capacity slightly overshoots the equipartition
value, oscillate back below the equipartion value (at a
temperature outside the scale in Fig.~\ref{bosCV}), and eventually
approach the limit from below. The curve marked COM shows the 
heat capacity when assuming that only the center of mass mode is 
excited. All the other curves will approach this at very large 
temperatures when the internal structure is washed out. However, 
as the plots clearly show, the approach is very different for 
different particle numbers and interaction strengths.

\begin{figure}
\includegraphics[width=0.48\textwidth]{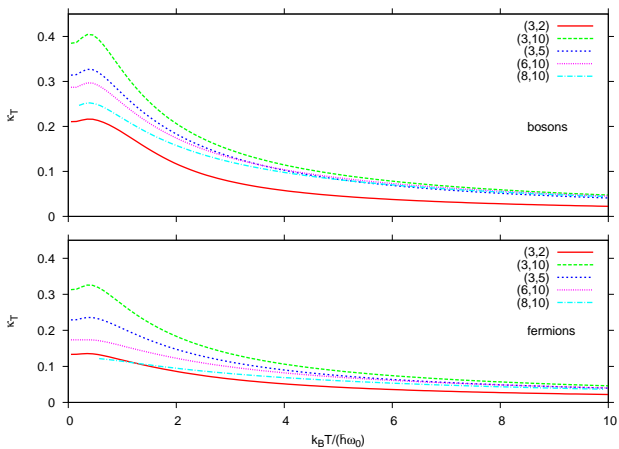}
\caption{Isothermal compressibility in two dimensions ($1/B_T$) divided by $N$
 for boson (upper) and fermions (lower) systems, identified by the number of
  particles and frequency ratios, $(N,\omega_r/\omega_0)$.  }
\label{bosKT}
\end{figure}

\begin{figure}
\includegraphics[width=0.48\textwidth]{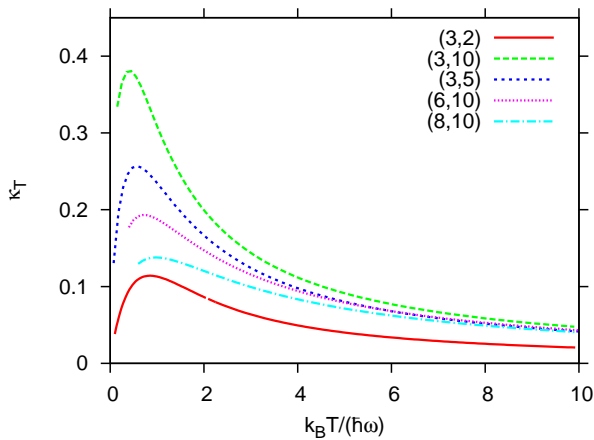}
\caption{ Isothermal compressibility ($1/B_T$) divided by $N$
  for fermionic systems in three dimensions, identified
  by the number of particles and frequency ratios,
  $(N,\omega_r/\omega_0)$. }
\label{3DferKT}
\end{figure}

In Fig.~\ref{bosKT}, we show the isothermal compressibility per
particle, $\kappa_T/N$, as defined in Eq.\eqref{e157}.
The high temperature behavior of a
harmonic system is $1/T$.
The compressiblity shows a small increase through a
maximum at very small temperature followed by steady decrease towards
zero at large temperature with a $1/T$ slope. 
The compressiblity indicates how easy
it is to squeeze the system and a large value indicates that the system
is very susceptible to compression. We see that the more strongly 
interacting systems (larger $\omega_r/\omega_0$) have larger
$\kappa_T$. This comes from the fact that the attraction in these
systems will make it energetically favourable to contract, due to
the larger degeneracy of the interaction frequency, $\omega_r$. This 
also explains why $\kappa_T$ increases with $N$.
Comparing bosons and fermions we find the same
qualitative behavior.  The fermions are slightly less compressible,
with a sharper dependence on particle number and interaction strength. 
This is a common feature of Fermi systems and usually attributed 
to the Pauli principle. For comparison,
Fig.~\ref{3DferKT} shows the compressibility for fermions in three
spatial dimensions, which shows the same qualitative 
behavior. The results for bosons are similar and we do not show them here. 
Notice that the peak features in the compressibility are sharper in 
the three-dimensional case.
Our system is similar to an attractively 
interacting Fermi gas where superfluidity is expected 
to show a signature in the compressibility \cite{ku2011}. Our results
are consistent with the fact that phase transitions are less pronounced
in lower dimensions.

\section{Summary and conclusions}\label{seccon}
The harmonic approximation is extremely useful because its simplicity allows
transparent calculations of otherwise complicated properties. The only
approximation lies in the choice of harmonic potentials acting on each
particle and between pairs of particles. One can therefore think
of the harmonic approximation as an effective interaction scheme
and subsequently investiagate the behavior of its predictions under
changes in the input parameters. The latter should naturally be
connected to whatever realistic physical system one is interested
in studying.

In this paper, we explore the harmonic approximation scheme by 
considering the Hamiltonian for a many-body system consisting
of a given number of identical particles. In particular, we 
explore the consequences of symmetry requirements on the 
properties of the system. This is done with thermodynamic 
applications in mind since this is a venue where the quantum
statistics plays a decisive role at low temperatures. We 
develop a new method for counting the number of correctly
symmetrized {\it many-body} states that reduces the complications
that arise from this fundamental problem of statistical mechanics
and thermodynamics. 

The advantage of the harmonic approach is obviously that 
the energy spectrum is analytically known.
However, the degeneracy of each many-body state of given total
energy still
remains to be determined before the partition function is fully
defined and possible to compute numerically.  
We design a novel
procedure to obtain the degeneracy of each state, subject to
requirements of symmetry and antisymmetry appropriate for bosons and
fermions, respectively.  We separate the completely permutation
symmetric center of mass motion from the relative motion, which then
has to carry the symmetry corresponding to bosons or fermions.  We
count by subtracting the number of states of different quanta in the
center of mass motion from the total number of non-interacting states
of given symmetry. 

To demonstrate the method, we consider the
case of a two-dimensional system with identical bosons or
fermions (with no internal degrees of freedom). Within the 
canonical ensemble, we compute
the partition function, and from it the free energy, entropy, 
heat capacity, and compressibility of the system. This is 
done for a relatively small number of particles (up to 20).
The method is a considerable improvement over the brute force
method where one explicitly checks for symmetry properties by
exchanging all pairs of particles one by one. However, it is
still computationally involved when going beyond the particle
numbers considered here. However, as is known from for instance
the virial expansion \cite{virial}, it is often enough to consider small 
particle numbers and then extrapolate to large system sizes 
from this information.

The effective harmonic interaction can be related to quantities
in realistic systems and we discuss a case of great usefulness 
within the realm of ultracold atomic gases, that of two 
particles interaction in a harmonic trap through a two-body
interaction of zero-range. While bosons can interact in the 
$s$-wave channel originally considered by Busch {\it et al.} \cite{busc98},
identical fermions must have an antisymmetric relative wave function
and thus an $s$-wave interaction of short-range will vanish. 
We therefore have to consider the $p$-wave channel, but we find
that the effective harmonic interaction frequencies are very similar
to the $s$-wave case in the two-dimensional setup that we consider 
here. Therefore we have chosen to parametrize the discussion of 
the thermodynamic quantities by the harmonic oscillator 
frequency of the two-body interaction itself. One can then 
make the connection to a realistic system by working backwards
through the model of Busch {\it et al.}.

Our numerical results show that the
low-temperature behavior reflects shell structure for the
particle numbers we study. The large-temperature equipartition limits
are recovered for energy and heat capacity, and
the transition from small
to large temperature is fastest for smaller two-body interaction strength.
The qualitative behavior is rather similar for bosons and fermions. However,
bosons always have a gap in the low-energy spectrum since one quantum
of excitation of the relative motion must be antisymmetric and
therefore forbidden. This leads to a slower variation with
temperature, since this gap has to be overcome before the number of
available states goes up. Our results for the density of states
turn out to scale with the number of relative excitation quanta
in a manner that is very similar to the treatment of the many-body density of
states for a uniform Fermi gas. Surprisingly, the bosonic many-body 
level density scales similarly to the fermionic one, although the 
particle number enters differently in our interpolation formulas.

In future studies it will be interesting to consider also multi-component
systems, something which is simply done within the harmonic approach
since the level counting can be factorized in the different
components. Also, a study of the virial expansion based on the 
harmonic approach is currently on-going, both in two and three
dimensions. In addition, we note that one dimensional system have 
attracted a lot of attention recently due to their realization 
in ultracold atomic physics \cite{bloch2008}. It would be
interesting to test our prediction against some of the models
that are being explored in the experiments and for which
a number of exactly solvable many-body models are known \cite{sutherland04}.
Furthermore, recent experiments studying ultracold few-body 
two-component Fermi systems (particle numbers of ten or less) \cite{jochim10}
would be an interesting comparison for the harmonic approximation.


\begin{thebibliography}{99}

\bibitem{magda00} M. A. Zaluska-Kotur, M. Gajda, A. Orlowski, and J. Mostowski, Phys. Rev. A {\bf 61}, 033613 (2000);
J. Yan, J. Stat. Phys. {\bf 113}, 623 (2003);
M. Gajda, Phys. Rev. A {\bf 73}, 023603 (2006);
\bibitem{arms11} J. R. Armstrong, N. T. Zinner, D. V. Fedorov, 
and A. S. Jensen, \textit{J. Phys. B: At. Mol. Opt. Phys}, \textbf{44} (2011)055303.

\bibitem{fetter71} A.~L. Fetter and J.~D. Walecka: \emph{Quantum Theory of Many-Particle Systems}, (McGraw-Hill, San Francisco, 1971).

\bibitem{bohr74} A. Bohr and B.~R. Mottelson: \emph{Nuclear Structure, Vol 1}, (Benjamin, New York, 1969).

\bibitem{sutherland04} B. Sutherland: \emph{Beautiful Models}, (World Scientific Publishing Co., Singapore, 2004); 
R.~J. Baxter: \emph{Exactly Solved Models in Statistical Mechanics}, (Academic Press, New York, 1982);
V.~E. Korepin: \emph{Exactly Solvable Models of Strongly Correlated Electrons}, (World Scientific Publishing Co., Singapore, 1994). 

\bibitem{bloch2008} I. Bloch, J. Dalibard, and W. Zwerger, Rev. Mod. Phys. {\bf 80}, 885 (2008).

\bibitem{chin10} C. Chin, R. Grimm, P.~S. Julienne, and E. Tiesinga, Rev. Mod. Phys. {\bf 82}, 1225 (2010).

\bibitem{jochim10} F. Serwane, G. Z{\"u}rn, T. Lompe, T.~B. Ottenstein, A.~N. Wenz, and S. Jochim, Science {\bf 332}, 6027 (2010);
G. Z{\"u}rn, F. Serwane, T. Lompe, A.~N. Wenz, M.~G. Ries, J.~E. Bohn, S. Jochim, arXiv:1111.2727v2.

\bibitem{bethe49} H.~A. Bethe, Phys. Rev. {\bf 76}, 38 (1949).

\bibitem{busc98} T. Busch, B.~G. Englert, K. Rz\c{a}\.zewski, and M. Wilkens, Found. Phys. {\bf 28}, 548 (1998).

\bibitem{stoferle06} T. St{\"o}ferle, H. Moritz, K. G{\"u}nter, M. K{\"o}hl, and T. Esslinger, Phys. Rev. Lett. {\bf 96}, 030401 (2006);
T. Volz {\it et al.}, Nature Phys. {\bf 2}, 692 (2006);
G. Thalhammer {\it et al.}, Phys. Rev. Lett. {\bf 96}, 050402 (2006);
C. Ospelkaus {\it et al.}, Phys. Rev. Lett. {\bf 97}, 120402 (2006).

\bibitem{haxton02} W.~C. Haxton and T. Luu, Phys. Rev. Lett. {\bf 89}, 182503 (2002). 
I. Stetcu, B.~R. Barrett, and U. van Kolck, Phys. Lett. B {\bf 653}, 358 (2007);
I. Stetcu, B.~R. Barrett, U. van Kolck, and J.~P. Vary, Phys. Rev. A {\bf 76}, 063613 (2007);
Y. Alhassid, G.~F. Bertsch, and L. Fang, Phys. Rev. Lett. {\bf 100}, 230401 (2008);
N.~T. Zinner, K. M{\o}lmer, C. {\"O}zen, D.~J. Dean, and K. Langanke, Phys. Rev. A {\bf 80}, 013613 (2009);
I. Stetcu, J. Rotureau, B.~R. Barrett, and U. van Kolck, Ann. Phys. {\bf 325}, 1644 (2010);
T. Luu, M.~J. Savage, A. Schwenk, and J.~P. Vary, Phys. Rev. C {\bf 82}, 034003 (2010);
J. Rotureau, I. Stetcu, B.~R. Barrett, M.~C. Birse, and U. van Kolck, Phys. Rev. A {\bf 82}, 032711 (2010).

\bibitem{hu2007} H. Hu , P.~D. Drummond, and X.-J. Liu, Nature Phys. {\bf 3}, 469 (2007).
\bibitem{ufermi} M. Horikoshi, S. Nakajima, M. Ueda, and T. Makaiyama, Science {\bf 327}, 442 (2010);
S. Nascimb{\'e}ne {\it et al.}, Nature {\bf 463}, 1057 (2010); N. Navon, S. Nascimb{\'e}ne, F. Chevy, and 
C. Salomon, Science {\bf 328}, 729 (2010); C. Cao {\it et al.}, Science {\bf 331}, 58 (2011).
\bibitem{fluc} C. Sanner {\it et al.}, Phys. Rev. Lett. {\bf 105}, 040402 (2010); 
T. M{\"u}ller {\it et al.}, Phys. Rev. Lett. {\bf 105}, 040401 (2010).
\bibitem{ku2011} M.~J.~H. Ku, A.~T. Sommer, L.~W. Clark, and M.~W. Zwierlein, arXiv:1110.3309v1.
\bibitem{hung2011} C.~L. Hung, X. Zhang, N. Gemelke, and C. Chin, Nature {\bf 470}, 236 (2011).

\bibitem{mcqu76} D.~A. McQuarrie: \emph{Statistical Mechanics}, (HarperCollins, New York, 1976).
\bibitem{huan87} K. Huang: \emph{Statistical Mechanics}, 2nd ed. (John Wiley \& Sons, New York, 1987).
\bibitem{wong98} S.~S.~M. Wong: \emph{Introductory Nuclear Physics}, (Wiley Interscience, New York, 1998).

\bibitem{lovelace87} R.~V.~E. Lovelace and T.~J. Tommila, Phys. Rev. A {\bf 35}, 3597 (1987);
G. Baym and C.~J. Pethick, Phys. Rev. Lett. {\bf 76}, 6 (1996).
\bibitem{fu03} H. Fu, Y. Wang, and B. Gao, Phys. Rev. A {\bf 67}, 053612 (2003);
N.~T. Zinner and M. Th{\o}gersen, Phys. Rev. A {\bf 80}, 023607 (2009);
M. Th{\o}gersen, N.~T. Zinner, and A.~S. Jensen, Phys. Rev. A {\bf 80}, 043625 (2009).

\bibitem{armstrong2010} J.~R. Armstrong, N.~T. Zinner, D.~V. Fedorov, and A.~S. Jensen, Europhys. Lett. {\bf 91}, 16001 (2010);
N.~T. Zinner, B. Wunsch, D. Pekker, and D.-W. Wang, arXiv:1009.2030v3;
N.~T. Zinner {\it et al.}, arXiv:1105.6264v1.

\bibitem{armstrong2011} J.~R. Armstrong, N.~T. Zinner, D.~V. Fedorov, and A.~S. Jensen, arXiv:1106.2102v1;
A.~G. Volosniev, D.~V. Fedorov, A.~S. Jensen, and N.~T. Zinner, arXiv:1109.4602v1.

\bibitem{volosniev2011}
A.~G. Volosniev {\it et al.}, J. Phys. B: At. Mol. Opt. Phys. {\bf 44}, 125301 (2011);
A.~G. Volosniev, D.~V. Fedorov, A.~S. Jensen, and N.~T. Zinner, Phys. Rev. Lett. {\bf 106}, 250401 (2011);
A.~G. Volosniev {\it et al.}, arXiv:1112.2541v1.

\bibitem{wunsch2011} B. Wunsch {\it et al.}, Phys. Rev. Lett. {\bf 107}, 073201 (2011); 
N.~T. Zinner {\it et al.}, Phys. Rev. A {\bf 84}, 063606 (2011).

\bibitem{pwave2011} N.~T. Zinner, arXiv:1111.1565v1.
\bibitem{kanjilal2006} K. Kanjilal and D. Blume, Phys. Rev. A {\bf 73}, 060701(R) (2006).
\bibitem{pri2006} L. Pricoupenko, Phys. Rev. A {\bf 73}, 012701 (2006).
 
\bibitem{virial} T.-L. Ho and E.~J. Mueller, Phys. Rev. Lett. {\bf 92}, 160404 (2004); X.-J. Liu, H. Hu, and P.~D. Drummond, 
Phys. Rev. Lett. {\bf 102}, 160401 (2009); Phys. Rev. A {\bf 82}, 023619 (2010); S.-G. Peng, S.-Q. Li, P.~D. Drummond, and 
X.-J. Liu, Phys. Rev. A {\bf 83}, 063618 (2011). 
\end{thebibliography}
\end{document}